\documentclass[aps,preprint,showpacs,showkeys]{revtex4}
\usepackage{amssymb}

\begin{document}

\title{Tsallis statistics and Langevin equation with multiplicative noise in
different orders of prescription}
\author{Kwok Sau Fa}
\affiliation{Departamento de F\'{\i}sica, Universidade Estadual de Maring\'{a}, Av.
Colombo 5790, 87020-900, \ Maring\'{a}-PR, Brazil, Tel.: 55 44 32614330,
Fax: 55 44 32634623}
\email{kwok@dfi.uem.br}

\begin{abstract}
Usually discussions on the question of interpretation in the Langevin
equation with multiplicative white noise are limited to the Ito and
Stratonovich prescriptions. In this work, a Langevin equation with
multiplicative white noise and its Fokker-Planck equation are considered.
From this Fokker-Planck equation a connection between the stationary
solution and the Tsallis distribution is obtained for different orders of
prescription in discretization rule for the stochastic integrals; the
Tsallis index $q$ and the prescription parameter $\lambda $ are determined
with the drift and diffusion coefficients. The result is quite general. For
application, one shows that the Tsallis distribution can be described by a
class of population growth models subject to the linear multiplicative white
noise.
\end{abstract}

\maketitle


The Langevin equation is a very important tool for describing out of
equilibrium systems \cite{risken,gardiner,coffey}. Moreover, this equation
has been extensively investigated; many properties and analytical solutions
of it have also been revealed. In particular, one considers the following
Langevin equation in one-dimensional space with a multiplicative white noise
term:

\begin{equation}
\frac{d\xi }{dt}=h(\xi )+g(\xi )\Gamma (t)\text{ ,}  \label{eq1}
\end{equation}%
where $\xi $ is a stochastic variable and $\Gamma (t)$ is the Langevin force
with the averages $\left\langle \Gamma (t)\right\rangle =0$ and $%
\left\langle \Gamma (t)\Gamma (\overline{t})\right\rangle =2\delta (t-%
\overline{t})$ \cite{risken}. $h(\xi )$ is the deterministic drift.
Physically, the additive noise ($g(\xi )$ constant) may represent the heat
bath acting on the particle of the system, and the multiplicative noise
term, for variable $g(\xi )$, may represent a fluctuating barrier. For $g=%
\sqrt{D}$ and $h(\xi )=0$, Eq. (\ref{eq1}) describes the Wiener process and
the corresponding probability distribution is described by a Gaussian
function. In the case of $g(\xi )$, some specific functions have been
employed to study, for instance, turbulent flows $(g(x)\sim \left\vert
x\right\vert ^{a})$ \cite{richard}. It should be noted that the linear
Fokker-Planck equation corresponding to the Langevin equation (\ref{eq1})
involves different orders of prescription in discretization rule for the
stochastic integrals \cite{risken}. The aim of this work is to obtain a
class of these Fokker-Planck equations in which their stationary solutions
are described by the Tsallis distribution \cite{tsallis1,tsallis2}. In
particular, the Tsallis distribution has been connected to a variety of
natural systems (see, for instance, \cite%
{plastino1,plastino2,lyra,beck,gellmann,zander}), and it is given by%
\begin{equation}
W_{q}(x)=N\left[ 1-\beta (1-q)U(x)\right] ^{\frac{1}{1-q}}\text{ ,}
\label{eq1a}
\end{equation}%
where $q$ is the Tsallis index and $U(x)$ is the potential. Notice that this
result extends the one given in Ref. \cite{borland}. Moreover, one shows
that the Tsallis distribution can be described by a class of population
growth models subject to the linear multiplicative white noise.

In one-dimensional space, the forward Fokker-Planck equation \cite%
{risken,hanggi,kwok1,kwok2} for the probability distribution corresponding
to the Langevin equation (\ref{eq1}) is given by

\begin{equation}
\frac{\partial W(x,t)}{\partial t}=-\frac{\partial }{\partial x}\left[
D_{1}(x)W(x,t)\right] +\frac{\partial ^{2}}{\partial x^{2}}\left[ D(x)W(x,t)%
\right] \text{ ,}  \label{eq2}
\end{equation}%
where $D_{1}(x,t)$ and $D(x,t)$ are the drift and diffusion coefficients
given by%
\begin{equation}
D_{1}(x)=h(x)+2\lambda \frac{dg(x)}{dx}g(x)  \label{eq2a}
\end{equation}%
and%
\begin{equation}
D(x)=g^{2}(x)\text{, }  \label{eq2b}
\end{equation}%
and $\lambda $ ($0\leq \lambda \leq 1$) is the prescription parameter due to
the discretization rule for the stochastic integrals. For instance, $\lambda
=0$ corresponds to the Ito prescription, $\lambda =1/2$ corresponds to the
Stratonovich prescription and $\lambda =1$ corresponds to the transport or H%
\"{a}nggi-Klimontovich prescription \cite{hanggi,kwok1,kwok2}. Notice that
Eq. (\ref{eq2a}) does not have a spurious drift only for the Ito
prescription, and the prescription parameter is directly linked to the
coefficient $g(x)$ of the fluctuation. The stationary solution of Eq. (\ref%
{eq2}) can be described as follows:%
\begin{equation}
W(x)=N\exp \left( \int \frac{dx}{D(x)}\left( D_{1}(x)-\frac{dD(x)}{dx}%
\right) \right) \text{ .}  \label{eq3}
\end{equation}%
Now this distribution will be equal to the Tsallis distribution (\ref{eq1a})
if the following relation is satisfied 
\begin{equation}
\frac{1}{D(x)}\left[ h(x)+\left( \lambda -1\right) \frac{dD(x)}{dx}\right] =-%
\frac{\beta }{1-\beta (1-q)U(x)}\frac{dU(x)}{dx}\text{ .}  \label{eq4}
\end{equation}

Note that this last relation is quite general, i.e, it is valid for a large
class of drift and diffusion coefficients and any prescription. In
particular, Eq. (\ref{eq4}) has been used to show that the probability
distribution in momentum space of the atom-laser interaction in the optical
lattice is described by the Tsallis distribution \cite{lutz}. For another
application of Eq. (\ref{eq4}), one considers the models of population
growth which are frequently described by nonlinear differential equations
without the independent variable (time) explicitly. For instance, the
classical Verhulst logistic equation is a simple nonlinear model of
population growth and it has been employed as a starting point to formulate
various generalized models \cite%
{montroll,renshaw,nisbet,davis,rubi,birch,wallace,peleg}. Moreover, this
logistic equation has been successfully used to model many laboratory
populations such as \ yeast growth in laboratory cultures, growth of the
Tasmanian and South Australian sheep populations \cite{renshaw} and
self-organization at macromolecular level \cite{eigen,jackson}. Besides, the
parameters involved in these models are subject to fluctuations and various
types of noises which may affect the replication processes. In this work,
one considers the following class of population growth models \cite{saka}: 
\begin{equation}
h(x)=r\frac{x\left[ 1-\left( \frac{x}{K}\right) ^{\nu }\right] }{\mu \left[
1-\left( 1-\frac{\nu }{\mu }\right) \left( \frac{x}{K}\right) ^{\nu }\right] 
}\text{ ,}  \label{eq10}
\end{equation}%
where $x(t)$ is the number of population alive at time $t$, $\mu $ and $\nu $
are real positive parameters, $r$ is the intrinsic growth rate and $K$ is
the carrying capacity. The deterministic model given by Eqs. (\ref{eq1}) and
(\ref{eq10}), with $g(x)=0$, contains the classical growth models such as
Verhulst logistic model ($\mu =1$ and $\nu =1$), Gompertz model ($\mu =0$
and $\nu \rightarrow 0$), Schoener model ($\mu =0$ and $\nu =1$), Richards
model ($\mu =0$ and $0\leqslant \nu <\infty $) and Smith model ($0\leqslant
\mu <\infty $ and $\nu =1$). In applications, the influence of the noise
(such as a change in temperature, food and water supplies) is usually
considered by the linear multiplicative noise of the effective birth rate 
\cite{zyga} which is given by $g(x)=\epsilon x$, where $\epsilon $ is
associated with the noise strength; in this case, the diffusion coefficient
is given by $D(x)=\epsilon ^{2}x^{2}$. For nonlinearly coupled noise see for
instance \cite{zyga2}. For the system described by Eqs. (\ref{eq1}) and (\ref%
{eq10}) without the influence of the noise ($g(x)=0$), the solution can be
determined implicitly and it is given by 
\begin{equation}
rt=\ln \left( \frac{\left( \frac{x}{K}\right) ^{\mu }\left( 1-\left( \frac{%
x_{0}}{K}\right) ^{\nu }\right) }{\left( \frac{x_{0}}{K}\right) ^{\mu
}\left( 1-\left( \frac{x}{K}\right) ^{\nu }\right) }\right) \text{ ,}
\label{eq10a}
\end{equation}%
where $x_{0}=x(t=0)$ is the initial value. Fig. 1 shows the evolution of
population for different models; they exhibit sigmoidal shapes.

Now, with the influence of the noise one considers the linearly coupled \
noise; from Eq. (\ref{eq3}) yields%
\begin{equation}
W(x)\sim x^{\frac{r}{\mu \epsilon ^{2}}+2\left( \lambda -1\right) }\left[
1-\left( 1-\frac{\nu }{\mu }\right) \left( \frac{x}{K}\right) ^{\nu }\right]
^{\frac{r}{\mu \epsilon ^{2}\left( \mu -\nu \right) }}\text{ .}  \label{eq11}
\end{equation}%
In particular, for $\mu =\nu $ one obtains 
\begin{equation}
W(x)\sim x^{\frac{r}{\mu \epsilon ^{2}}+2\left( \lambda -1\right) }\exp
\left( -\frac{r}{\mu ^{2}\epsilon ^{2}}\left( \frac{x}{K}\right) ^{\mu
}\right) \text{ .}  \label{eq11a}
\end{equation}%
The distribution (\ref{eq11a}) is well-known \ and it is called the Weibull
distribution. Besides, the distribution (\ref{eq11}) presents interesting
aspects. \ In order to retain a consistent probabilistic interpretation, the
cut-off condition imposes $W(x)=0$ whenever $\mu >\nu $ and $\left[ 1-\left(
1-\frac{\nu }{\mu }\right) \left( \frac{x}{K}\right) ^{\nu }\right] <0$. \
For $\lambda =1$ and \ positive values for $r$ and $\mu $ the distribution (%
\ref{eq11}) is zero at the origin; in this case the populations can always
survive. However, for $\lambda \neq 1$ the distribution (\ref{eq11}) is
divergent at the origin for 
\begin{equation}
-1<\frac{r}{\mu \epsilon ^{2}}+2\left( \lambda -1\right) <0\text{ ;}
\label{eq11aa}
\end{equation}%
this means that the order of prescription in the discretization rule for the
stochastic integrals can prevent population extinction.

From the solution (\ref{eq11}) the Tsallis distribution can be obtained by
setting%
\begin{equation}
\frac{r}{\mu \epsilon ^{2}}+2\left( \lambda -1\right) =0\text{, \ \ \ }%
\lambda \neq 1  \label{eq11b}
\end{equation}%
and 
\begin{equation}
1-q=\frac{\mu \epsilon ^{2}\left( \mu -\nu \right) }{r}\text{ .}
\label{eq11c}
\end{equation}%
Eq. (\ref{eq11b}) implies that%
\begin{equation}
\frac{r}{2\mu \epsilon ^{2}}<1  \label{eq11da}
\end{equation}%
and 
\begin{equation}
\lambda =1-\frac{r}{2\mu \epsilon ^{2}}\text{ .}  \label{eq11d}
\end{equation}%
The expression (\ref{eq11c}) shows that the Tsallis index $q$\ can assume
any real value, and it is also associated with the microscopic parameters of
the system. For $q=1$ yields $\mu =\nu $, and the distribution (\ref{eq11})
reduces to the stretched exponential for $0<\mu <1$, exponential function
for $\mu =1$ and compressed exponential for $\mu >1$. It is worth noting
that the distribution (\ref{eq11}) with the conditions (\ref{eq11b}) and (%
\ref{eq11c}) also satisfy the relation (\ref{eq4}) for%
\begin{equation}
U(x)=x^{\nu }  \label{eq12a}
\end{equation}%
and 
\begin{equation}
\beta =\frac{r}{\mu ^{2}\epsilon ^{2}K^{\nu }}\text{ .}  \label{eq12b}
\end{equation}

In summary, a Langevin equation with multiplicative white noise and its
corresponding linear Fokker-Planck equation for different orders of
prescription in discretization rule for the stochastic integrals have been
considered. In particular, the stationary solution of the Fokker-Planck
equation has been connected with the Tsallis distribution for \ generic
drift and diffusion coefficients. As an application one showed that the
stationary solution of a class of population growth models subject to the
linear multiplicative white noise could be described by the Tsallis
distribution for different orders of prescription. In these models, the
Tsallis index $q$ has been connected with the microscopic parameters.
Interesting aspects have been achieved; the order of prescription in
discretization rule for the stochastic integrals plays a key role to connect
with the Tsallis distribution. Moreover, the change in the prescription
parameter $\lambda $ of a given system may lead to the extinction or
survival of a population. As is expected, the prescription \ parameter can
modify the behavior of the system considerably, i.e, different prescription
may describe different behavior.

The author acknowledges partial financial support from the Conselho Nacional
de Desenvolvimento Cient\'{\i}fico e Tecnol\'{o}gico (CNPq), a Brazilian
agency.

\newpage

\begin{center}
\textbf{Figure Captions}

\bigskip

\bigskip

\bigskip
\end{center}

Fig. 1 - \ Plots of the evolution of population for different growth models
described by Eq. (\ref{eq10a}), in arbitrary units. The parameter values are
given by $r=0.5$, $x_{0}=0.2$ \ and $K=10.$

\end{document}